\documentstyle[11pt]{article}

\font\uno=cmcsc10 scaled 1200
\font\dos=cmti10 scaled 1200

\newcommand{\be}{\begin{equation}}
\newcommand{\ee}{\end{equation}}
\newcommand{\bea}{\begin{eqnarray}}
\newcommand{\beas}{\begin{eqnarray*}}
\newcommand{\eea}{\end{eqnarray}}
\newcommand{\eeas}{\end{eqnarray*}} 
\newcommand{\ba}{\begin{array}}
\newcommand{\ea}{\end{array}}
\newcommand{\bi}{\begin{itemize}}
\newcommand{\ei}{\end{itemize}}
\newcommand{\ben}{\begin{enumerate}}
\newcommand{\een}{\end{enumerate}}
\newcommand{\bt}{\begin{tabular}}
\newcommand{\et}{\end{tabular}}
\renewcommand{\theequation}{\arabic{section}.\arabic{equation}}
\newcommand{\eq}{\setcounter{equation}{0}}

\newcommand{\ov}{\overline}
\newcommand{\la}{\langle}
\newcommand{\ra}{\rangle}
\def\harr#1#2{\smash{\mathop{\hbox to .5in{\rightarrowfill}}
\limits^{\scriptstyle#1}_{\scriptstyle#2}}}

\def\harrl#1#2{\smash{\mathop{\hbox to .5in{\leftarrowfill}}
\limits^{\scriptstyle#1}_{\scriptstyle#2}}}

\def\refname{\uno References} 
\def\thebibliography#1{\section*{\refname}
\list 
 {\arabic{enumi}.}{\settowidth\labelwidth{[#1]}\leftmargin\labelwidth 
 \advance\leftmargin\labelsep 
 \usecounter{enumi}} 
 \def\newblock{\hskip .11em plus .33em minus .07em} 
 \sloppy\clubpenalty4000\widowpenalty4000 
 \sfcode`\.=1000\relax} 
 
\begin{document}

\title{Symmetry Breaking in $[SU(6)]^3\times Z_3$}
\author{
A. P\'erez-Lorenzana$^{a}$, D. E. Jaramillo$^{a,b}$, William A.
Ponce$^{b}$, and Arnulfo Zepeda$^{a}$\\
{\dos a Departamento de F\'{\i}sica,}\\
{\dos Centro de Investigaci\'on y de Estudios Avanzados del I.P.N.}\\
{\dos Apdo. Post. 14-740, 07000, M\'exico, D.F., M\'exico.}\\
{\dos b Departamento de F\'\i sica, Universidad de Antioquia, 
A.A. 1226, Medell\'\i n, Colombia.}}

\maketitle
{{\uno Abstract}.\small
 We analize the different ways for the spontaneous breaking of the gauge
symmetry, for the $[SU(6)]^3\otimes Z_3$ family unification model.  In
particular we study the consequences of a previous selection for the
vacuum expectation values of the Higgs fields, showing that such set
predicts unwanted flavor changing neutral currents at the $m_Z=91 GeVs$
mass scale. A new set of vacuum expectation values which solves this
problem is proposed. \\[2ex]
 }

\newpage

\section{\uno Introduction.}

Although the Standard Model (SM) is a successful theory which is in good
agreement with the experimental results~\cite{pdg}, it leaves several
primordial aspects unanswered. Outstanding among them is the so called
flavor problem which is the lack of predictions for the fermion mass
spectrum, the number of families in nature and the small values for the
quark mixing angles. 
In order to get an answer to this problem we believe that there is a
more fundamental theory, not far away from the present experimental
energies. This is one of the motivation for Grand Unified  
Theories (GUTs)~\cite{guts} which
are extensions of the SM gauge structure $SU(3)_c\otimes SU(2)_L\otimes
U(1)_Y$, into larger groups with a single gauge coupling constant.

In Ref.~\cite{2} it was presented a variant of the three family extension 
of the  Pati-Salam\cite{ps} model which does not have mirror fermions and is 
renormalizable. In this model the known families belong to a single
irreducible  representation of the local gauge group $[SU(6)]^3\times Z_3$,
each family being defined by the dynamics of the left-right 
symmetric extension (LRSE)~\cite{moha} 
$SU(3)_c\otimes SU(2)_R\otimes SU(2)_L\otimes U(1)_{(B-L)}$ of the SM.

In Ref.~\cite{tuning}  the new model  was systematically studied, paying
special attention to its particle content, the symmetry  breaking (SB)
pattern, the mass
scales, the free Lagrangean for all the gauge bosons (including their mass
terms), the bare masses for all the exotic fermion fields in  the model, and
the interacting Lagrangean with all the known and predicted  gauge
interactions. Even though the results  presented in Ref.~\cite{tuning} are
important for the analysis of the mass spectrum  of the known quarks
and leptons in the context of the model, there are 
several problems in the SB scheme proposed, because the set of Higgs
fields and vacuum expectation values (vevs)
used do not break the local symmetry $[SU(6)]^3\times Z_3$
down to the SM symmetry. As a matter of fact, an extra $U(1)$
symmetry is predicted by this SB with a nonuniversal coupling to the
standard matter at the $m_Z$ scale. Moreover, the spontaneous 
breaking of this symmetry mixes the $Z$ gauge field of the SM with
the field associated to the extra symmetry, giving
Flavor Changing Neutral Currents  (FCNC) mediated by a gauge boson field
with a mass of the order of $m_Z$. Since this neutral currents are not
allowed by the low energy experimental results, a careful analysis of the
SB pattern for the $[SU(6)]^3\times Z_3$ is needed. That is the aim of the
work presented in this paper.

The paper is organized in the following way: in the next section we
briefly review the model $[SU(6)]^3\times Z_3$. In the first part of
section 3 we analyze general
ways for implementing the spontaneous SB in the context of the model, in
the second part of section 3 we 
discuss the problem with  the SB scheme used in ref.~\cite{tuning}, 
and we propose a new pattern which solves the puzzle. The 
renormalization group equation analysis for the new SB scheme is presented
in section 4 and the mass scales for the new model are
estimated. We write our conclusions and some comments in the last
section. An appendix with technical information is included at the end.

\section{\uno The model.}
\eq

The model under consideration is based on the local gauge group 
\be  G \equiv
SU(6)_L\otimes SU(6)_c\otimes SU(6)_R\times Z_3 
\ee  
and unifies non-gravitational forces with transitions among
three families.  In Eq. (2.1) $\otimes$ indicates a direct product,
$\times$ a semidirect one, and $Z_3$ is a three-element cyclic group
acting upon $[SU(6)]^3$ such that if $(A,B,C)$ is a representation of
$[SU(6)]^3$ with $A$ a representation of the first factor, $B$ of the
second and $C$ of the third, then $Z_3(A,B,C)\equiv (A,B,C) \oplus (B,C,A)
\oplus (C,A,B)$ is an irreducible representation (irrep) of $G$. $SU(6)_c$
is a vector-like
group which includes three hadronic and three leptonic colors, and
has as a subgroup the $SU(3)_c \otimes U(1)_{B-L}$ group of the
LRSE model.
$SU(6)_L\otimes SU(6)_R$ includes the $SU(2)_L\otimes SU(2)_R$ gauge group
of the LRSE model. Among the special properties of this model we may
recall
that its gauge group, $G$, is the most economical unifying group for three 
families, with left-right symmetry and with (extended) vector color; 
it leads to 
(perturbative) stability of the proton~\cite{mass}. Furthermore, all the known
elementary fermions belong to an irrep
of $G$. On the other hand the presence of the horizontal group in 
$SU(6)_L\otimes SU(6)_R$ allows for the possibility of obtaining
predictions for the fermion mass spectrum \cite{2,tuning}.

\vskip.3cm

The 105 gauge fields (GFs) in $G$ can be divided into two sets: 70 of them
belonging to $SU(6)_L\otimes SU(6)_R$ and 35 being associated with $SU(6)_c$. 
The first set includes $W^\pm_L$ and $W^0_L$ (the GFs of $SU(2)_L$ in the
SM), the GFs associated with $SU(2)_R$; the 
GFs of the horizontal interactions, and new GFs of nonuniversal 
charged and neutral interactions. All of them have electric 
charges $0$ or
$\pm 1$. The generators of $SU(6)_{L(R)}$ may be written in a 
$SU(2)_{L(R)}\otimes SU(3)_{HL(HR)}$ basis as 
\be
\sigma_i\otimes I_3/2\sqrt{3},\qquad I_2\otimes\lambda_\alpha/2\sqrt{2}, \qquad
\sigma_i\otimes\lambda_\alpha/2\sqrt{2},    \label{LRgenerators}
\ee
where $\sigma_i, i=1,2,3$ are the three $2\times 2$ Pauli matrices,
$\lambda_\alpha, \alpha=1,...8$ are the eight $3\times 3$ Gell-Mann
matrices, and $I_2$ and $I_3$ are 
$2\times 2$ and $3\times 3$ identity matrices respectively.

The second set of gauge fields includes the eight gluon fields of
$SU(3)_c$, nine
lepto-quark GFs ($X_i$, $Y_i$ and $Z_i$, $i=1,2,3$, with electric
charges $-2/3$, $1/3$ and $-2/3$ respectively), their nine
charge conjugated fields, six
dilepton GFs ($P^\pm_a$, $P^0$ and $\tilde P^0$, $a=1,2$, with electric 
charges as indicated), and the GFs associated with diagonal generators in
$SU(6)_c$ and not taken into account already in $SU(3)_c$, including among
them the gauge field associated with the $(B-L)$ abelian generator of the
LRSE model.

\vskip.3cm

The fermion fields of the model are in the irrep
\be
\psi(108)_L = Z_3\psi (6,1,\ov 6)_L =\psi(6,1,\ov 6)_L \oplus 
\psi(\ov 6,6,1)_L \oplus \psi(1,\ov 6,6)_L, \label{psi}
\ee
with quantum numbers with respect to the SM factors $(SU(3)_c, SU(2)_L,
U(1)_Y)$ given by
\beas
\psi(\ov 6,6,1)_L &\equiv& \psi^\alpha_a:  \ \ \ \ \ 3(3,2,1/3) \oplus 
6(1,2,-1) \oplus 3(1,2,1),\\
\psi(1,\ov 6,6)_L &\equiv& \psi^A_\alpha:  \ \ \ \ \ 3(\ov 3,1,-4/3) \oplus 
3(\ov 3,1,2/3)\oplus 6(1,1,2) \oplus 9(1,1,0) \oplus 3(1,1,-2),\\
\psi(6,1,\ov 6)_L & \equiv& \psi^a_A:  \ \ \ \ \ 9(1,2,1) \oplus 9(1,2,-1),
\eeas
where $a,b,\dots,A,B,\dots,\alpha,\beta,\dots=1,\dots, 6$ label $L$,
$R$ and $c$ tensor indices, respectively. The known fermions are 
contained in $\psi(\ov 6,6,1)_L\oplus\psi(1,\ov{6},6)_L\subset \psi(108)$.

The electric charge operator in the model is given by
\be
Q = T_{ZL} + Y/2
\ee
where the hypercharge $Y/2 = T_{ZR} +{1\over 2} Y_{(B-L)}$ and 
$T_{ZL,R} =diag\{ 1,-1,1,-1,1,-1\}/2$ and $Y_{(B-L)} = 
diag\{1/3,1/3,1/3,-1,1,-1\}$ which act on the subspaces of the
fundamental irreps of $SU(6)_{L(R)}$ and $SU(6)_c$ respectively.

\section{\uno The spontaneous symmetry breaking pattern.}
\eq
\subsection{\uno General analysis.}

In order to achieve the spontaneous SB we introduce
appropriate Higgs scalars. Using the branching rules
\be \ba{rcl}
SU(6)_{L(R)}& \rightarrow &
SU(2)_{L(R)}\otimes SU(3)_{HL(R)} \\
 6 & \rightarrow & (2,3)\\
 15 & \rightarrow & (1,6) + (3, \ov{3} )\\
 21 & \rightarrow & (1,\ov{3}) + (3,6) \\
 35 & \rightarrow & (3,8)+(3,1)+(1,8) 
 \ea\quad
\ba{rcl}
SU(6)_c& \rightarrow & SU(3)_c\\
 6  & \rightarrow & (3) + 3 (1)\\
 15 & \rightarrow & (\ov {3}) + 3(3)+ 3(1) \\
 21 & \rightarrow & (6) + 6(1) + 3(3) \\
 35 & \rightarrow & (8)+_3(3)+3(\ov{3})+9(1), 
\ea \label{su2su3}
\ee
we can see that the vacuum expectation values (vevs) of a 6 of 
$SU(6)_L$ necessarily
break $SU(2)_L$; besides a Higgs field $\phi(18) = Z_3 \phi(6,1,1)$ is not 
sufficient to give tree-level masses to ordinary fermion fields. We 
therefore assume, as it was done in Ref.\cite{tuning}, that the last step 
of the SB chain is due to the vevs of a Higgs field $\phi_4 = \phi(108) =
Z_3 \phi(1,\ov{6},6)$, 
and that these vevs lie only in the electrically neutral 
directions of the $SU(6)_L \otimes SU(6)_R$ subspace, in such a way that
the {\it modified horizontal survival hypothesis}\cite{2,tuning} holds
(which states that at tree level the top quark is the only
standard matter fermion field acquiring mass, with
the masses of the other known fermions being generated as radiative
corrections). 

In order to comply with the {\it survival hypothesis}
(which states that when a gauge group $G$ is broken down to
$G_1\subset G$ at a mass scale $M_1$, all the fermion fields
belonging to real representations of $G_1$ must acquire
masses of order $M_1$\cite{georgi}) we demand
that the first steps of the  SB chain arise from vevs of Higgs fields of 
the type $Z_3\phi(\ov {n},1,n)$, where $n$ may be $15$ or $21$. 
For this kind of fields  their vevs have the general form
\be
\la\phi\ra = m\left[ 
\la\phi_{Lc}\ra \oplus \la\phi_{cR}\ra \oplus 
 \la\phi_{LR}\ra \right], \label{gvevs}
 \ee
 where the subindices indicate the subspaces involved in each term and $m$
 is the mass scale of the breaking implement by $\la\phi\ra$. The covariant
 derivative acting on a representation of the form 
$a_L\otimes a_c\otimes a_R$ of $[SU(6)]^3$, with $a_L$ ($a_c$ or $a_R$) 
a fundamental irrep of the factor $SU(6)_L$ ($SU(6)_c$ or $SU(6)_R$) is
 \be 
 {\cal D} = D_L\otimes {\bf 1}\otimes{\bf 1}\ \oplus \ 
 {\bf 1}\otimes D_c\otimes {\bf 1}\  \oplus\ {\bf 1}\otimes{\bf 1}\otimes D_R,
 \ee
being $D_i$ ($i = L,c,R$) the corresponding covariant derivative on the irrep
$a_i$ defined by $D_i^\mu = \partial^\mu + ig{\bf A}_i^\mu$, with ${\bf
A}_i^\mu = {1\over 2}\lambda_i^a A_{a,i}^\mu$ $a = 1, \dots, 35$,
where $\lambda_i^a$ are the
generators of $SU(6)_i$ normalized to $Tr\lambda^a_i\lambda^b_i =
2\delta^{ab}$. Also $A_{a,i}^\mu$ are the gauge bosons associated to the
generators $\lambda^a_i$ and $g$ is the gauge coupling constant of $G$.
For fields $\Phi$ in irreps 15 or 21 of an $SU(6)$ factor, the action of
the covariant derivative is
\be
 D_i^\mu\left(\Phi\right) = \partial^\mu\Phi + ig\left[{\bf
A}_i^\mu\Phi + \Phi{\bf A}^{\mu,T}_i\right]
\ee
where the last equation is stated in a $6\times 6$ matrix form. 

The mass Lagrangean for the gauge bosons produced by $\la\phi\ra$ is of
the form 
\[{\cal L}_{mass} = Tr[{\cal
D}(\la\phi\ra)]^\dagger[{\cal D}(\la\phi\ra)] =  {\cal L} _{Lc} + {\cal L}_{cR}
+ {\cal L}_{LR},\] 
where ${\cal L}_{Lc,cR,LR}$ are the corresponding contributions to the
Lagrangean by $\la\phi_{Lc,cR,LR}\ra$ respectively. They may be written as 
\be
{\cal L}_{ij} = 2 g^2Tr\bigg[\la\phi_{ij}\ra {{\bf A}_i}^2 
\la\phi_{ij}\ra \, +
\, \la\phi_{ij}\ra {{\bf A}_i}\la\phi_{ij}\ra {{\bf A}_i}^T\, +
\, (i\rightarrow j)\, 
- \, 4  \la\phi_{ij}\ra {{\bf A}_i}
 \la\phi_{ij}\ra{{\bf A}_j} \bigg], \label{lagmas}
\ee
with $ij = Lc,cR,LR$. While the first terms in (\ref{lagmas}) implement the
spontaneous breaking of the corresponding factor of $SU(6)_i$, giving
masses to the 
associated bosons, the last term mixes the bosonic fields of both sectors
involved, in such a way that the breaking of $SU(6)_i\otimes SU(6)_j$ via
$\la\phi_{ij}\ra$ is of the form 
$SU(6)_i\otimes SU(6)_j\rightarrow G_i\otimes G_j\otimes G_{mix}$
where the specific groups $G_{i(j)}$ depend only on the particular
direction of
the vevs in the $i$ ($j$) subspace, but the mixing symmetry is given by
the combined action of directions in both subspaces. 

According to the branching rules stated in (\ref{su2su3}), there are six 
$SU(2)_{L,R}$ singlets in 15 of $SU(6)_{L(R)}$ and three in irrep 21;
they are along the directions
\be \begin{array} {c c l}
  15 &:\qquad  & [1,4]-[2,3], \ \ [1,6]-[2,5], \ \ [3,6]-[4,5],\\
     &         & [1,2] ,\ \ [3,4], \ \ [5,6],  \\
  21 &:\qquad  & \{1,4\}-\{2,3\},\ \ \{1,6\}-\{2,5\},\ \ \{3,6\}-\{4,5\}. 
\end{array} \label{sing}
\ee
\noindent
where the notation is such that $[a,b]=ab-ba$, and $\{a,b\}=ab+ba$. 
The analysis shows~\cite{tesis} that if the Higgs fields get vevs along
these direction in the subspaces $L$ or $R$, the corresponding 
$SU(6)$ factor breaks down to the following subgroups: 

\be
SU(6)\left\{ \ba{c l}
\harr{[1,6]-[2,5]+[3,4]}{} & Sp(6)  \\
                           &        \\
\harr{[1,2]}{}             & SU(4)\otimes SU(2)\ea\right. \mbox{ and }
SU(6)\left\{ \ba{c l}
\harr{[1,4]-[2,3]}{}       & Sp(4)\otimes SU(2)\\
                           & \\
\harr{ \{1,4\}-\{2,3\}}{}  & [SU(2)]^3 \ea\right. \label{sbsin}
\ee
where we have written just the isomorphic residual symmetry group, with  
the specific structure of each subgroup depending on the particular
direction for the vevs. Other combinations of the vevs above produce 
similar SB patterns; for example, $[1,6]-[2,5]-[3,4]$ also breaks $SU(6)$
down to $Sp(6)$, $[3,4]$ or $[5,6]$ (instead of $[1,2]$) break $SU(6)$
down to $SU(4)\otimes SU(2)$, $[1,6]-[2,5]$ or $[3,6]-[4,5]$ (instead of  
$[1,4]-[2,3]$) break $SU(6)$ down to $Sp(4)\otimes SU(2)$, 
$\{1,6\}-\{2,5\}$ or $\{3,6\}-\{4,5\}$ instead of $\{1,4\}-\{2,3\}$  break
$SU(6)$ down to $[SU(2)]^3$, etc..

The reason for choosing these channels for the first step of the SB is
due to the following facts: they contain the $SU(2)_{L(R)}$ structures
of the LRSE group as subgroups; the vevs in the directions of the
singlets in irreps 15 and 21 assure an unbroken $SU(2)_{L(R)}$ factor;
and finally they comply properly with the survival hypothesis.

In order to break $SU(2)_R$, the only constraint on the vevs directions 
come from the demand that the generator associated to the hypercharge $Y$
must not be broken before the last step of the SB chain. In other words,
the vevs of the Higgs fields $\phi_3$ which breaks $SU(2)_R$ must satisfy 
\be 
Y\left(\la\phi_3\ra\right) = 0, \label{yp}
\ee
where again we choose $\phi_3$ of the form $Z_3\phi_3(\overline{n},1,n)$
with $n=15,21$; so it is also of the form given by (\ref{gvevs}). 
The simplest way to achieve the constraint (\ref{yp}) is imposing that
\be
T_{iR}\left(\la\phi_{3LR}\ra\right) = 0, \hspace{1cm} 
T_{iR}\left(\la\phi_{3Lc}\ra\right) = 0
\ee
and 
\be
T_{iR}\left(\la\phi_{3cR}\ra\right) \neq 0.
\ee
where $T_{iR}, i=1,2,3$ are the three generators for $SU(2)_R$. 
These constraints are achieved in general for the 
combinations $(\alpha,\beta)$ odd-odd (even-even) only if $(A, B)$ are
odd-odd (even-even). In the subspace $c$ there are just 3 directions in 21
with even-even indices
which are $\{4,4\},~\{4,6\}$ and $\{6,6\}$; while there is just one in
15 which is $[4,6]$. For the odd-odd combination there is only one
possibility, the $\{5,5\}$ in irrep 21.

For completeness let us mention finally that any of the diagonal
directions $\{i,i\}$, $i=1\dots 6$ in 21 break $SU(6)$ down $SU(5)$, while
other directions implement the breakings
\be
SU(6)\,
\harr{[2,4],[2,6],[4,6]}{}\, SU(4)\otimes SU(2)\quad \mbox{ and }
\quad  SU(6)\, 
 \harr{\{1,3\},\{1,5\},\{3,5\}}{\{2,4\},\{2,6\},\{4,6\}}  
 \, SU(4)\otimes U(1). \label{sbycr}
\ee

\subsection{\uno Problems with the old SB scheme.}

In a previous paper~\cite{tuning} the SB for this model was implemented by 
introducing four different sets of scalar fields: 
\[ \phi_1 = \phi(675) = Z_3\phi(\ov{15},1,15)\]
with vevs at the scale $M$ in the directions $[a,b],[A,B]= [1,6]=-[2,5]=-[3,4]$
and $[\alpha,\beta]=[5,6]$;
\[ \phi_2 = \phi(1323) = Z_3\phi(\ov{21},1,21)\]
with vevs at the scale $M'$ in the directions $\{a,b\},\{A,B\}=
\{1,4\}= -\{2,3\}$ and $[\alpha,\beta]=\{4,5\}$;
\[ \phi_3 = \phi(675) = Z_3\phi(\ov{15},1,15)\]
such that $\la\phi^{[a,b]}_{3[A,B]}\ra$ =
$\la\phi^{[\alpha,\beta]}_{3[a,b]}\ra = 0$, and 
$\la\phi^{[A,B]}_{3[\alpha,\beta]}\ra\neq 0$ with vevs at the scale $M_R$
in the
directions $[\alpha,\beta],[A,B]= [4,6]$.

The last step of the SB which breaks the SM symmetry was implemented by 
introducing the scalar fields 
\[ \phi_4 = \phi(108) = Z_3\phi(6,1,\ov{6})\]
with vevs $\la\phi^\alpha_{4A}\ra = \la\phi_{4\alpha}^a\ra =
 0$ and $\la\phi_{4a}^A\ra = m_Z$ for $A,a = 2,4,6$. 
 As it was shown in Refs.\cite{tuning} and \cite{mass}, the model
 with only two different mass scales $M_G$ and $m_Z$, such that
 \[
  G\stackrel{M_G}{\longrightarrow}SU(2)_L\otimes SU(3)_c\otimes U(1)_Y 
  \stackrel{m_Z}{\longrightarrow} SU(3)_c\otimes U(1)_{em}
 \]
is excluded by the analysis of the renormalization group equations, 
and the experimental values for $\alpha_i(m_Z), i=1,2,3$. The breaking
pattern with three
different mass scales, where the first step is $G\rightarrow 
SU(3)_c\otimes SU(2)_R\otimes SU(2)_L\otimes U(1)_{(B-L)}$ is also
forbidden\cite{tuning}. Then the hierarchy $M_R>>M_H\equiv M\sim M' >>
m_Z$ is suggested, and therefore the first step of the SB should
be implemented via $\la\phi_3\ra$. From the analysis presented in
Eq.(\ref{sbycr}) we note that
$\la\phi_3\ra$ breaks 
$SU(6)_L\otimes SU(6)_c\otimes SU(6)_R$ down to $G_M\equiv SU(6)_L\otimes
SU(4)_c\otimes
SU(2)_c\otimes SU(4)_R\otimes SU(2)_R\otimes U(1)_\Sigma$, where $U(1)_\Sigma$
is an abelian symmetry generated as a mixing in the $c$-$R$ subspaces,
where 
\be
\Sigma = {1\over \sqrt{2}} \left( \Sigma_c + \Sigma_R\right) 
\ee
with
\be
\Sigma_{c(R)} = diag\big(1,1,1,-2,1,-2\big)/\sqrt{6}.
\ee
The algebra shows that $\Sigma\la\phi_3\ra = 0$, even
though 
$\Sigma_{c(R)}\la\phi_3\ra\neq 0$. In the unbroken group the factor
$SU(4)_{c(R)}$ ( $SU(2)_{c(R)}$ ) acts on the indices $1,2,3,5$ ( $4,6$ ) of
the fundamental irrep of $SU(6)_{c(R)}$. 

The next step of the SB was implemented in Ref.\cite{tuning} by $\la
\phi_1
+ \phi_2 \ra$. The 
analysis of the bosonic mass Lagrangean~\cite{tesis} shows now that 
$G_M\rightarrow
SU(3)_c\otimes SU(2)_R\otimes SU(2)_L\otimes U(1)_{(B-L)}\otimes U(1)'$; 
where we can notice that, unlike the statement in  reference
\cite{tuning}, 
$\la\phi_1+\phi_2+\phi_3\ra$ does not break $G$ down the SM gauge group.
As a matter of fact, there is  an extra Abelian symmetry at the $m_z$
scale. The gauge boson associated with this $U(1)^\prime$ symmetry is 
\be
B' = \left( 9\sqrt{3} B_{(B-L)} - 15 \sqrt{6} W^0_R -
28 \sqrt{5} B_{Y'} + 140 H_L
+ 140 H_R\right)/10\sqrt{469},
\ee
where the fields involved are the  gauge bosons associated to the
generators 
$Y_{(B-L)}$,
$T_{ZR}$, 
$Y' = \ diag(1,1,1, -3,-2,2)/\sqrt{10}$, and $T_{HL,HR} =  
 diag(1,1,0,0,-1,-1)/\sqrt{2}$. 
 As it is easy to check, the generator $T'$ of $U(1)'$ satisfies
 $T'\la\phi_1\ra = T'\la\phi_2\ra = T'\la\phi_3\ra = 0$, but
$T'\la\phi_4\ra\neq 0$. Then because $Q\la\phi_4\ra = 0$, the symmetry is
properly broken at the $m_Z$ scale, predicting the correct low energy
unbroken symmetry
$SU(3)_c\otimes U(1)_{em}$. Nevertheless, since $T_{ZL}\la\phi_4\ra\neq
0$, the last step of the SB mixes $B'$ with the $Z$ standard field
producing 
two new neutral fields $Z_1$ and $Z_2$ of the form
 \be
 \left( \ba{c} Z_1\\Z_2\ea\right) = \left(\ba {cc} \cos\varepsilon &
 -\sin\varepsilon\\
 \sin\varepsilon & \cos \varepsilon\ea \right)\left(\ba{c} B'\\ Z\ea \right).
 \ee
 
Considering the mass Lagrangean produced by $\la\phi_4\ra$, decoupling
all
the fields with high masses, and introducing explicitly the $Z$ standard and the
photon fields, we obtain the mixing terms
\be
{\cal L}_{\la\phi_4\ra} = {3\over 28} g^2 M_Z^2\left\{ {811\over 134} {B'}^2 +
3\sqrt{69\over 67}\, B'\, Z  + {23\over 2} Z^2 \, \right\} + \dots.
\ee 
From here it is simple to compute the mixing angle 
$\sin\varepsilon$ which is 
\be 
\sin\varepsilon = {3\sqrt{67}\sqrt{69}\over 2
(223)^{1/4}\sqrt{14}\sqrt{\alpha_0}}\simeq 0.2520.
\ee
 where $\alpha_0 = 365 + 28\sqrt{223}$. Then the mixing is large and its
effects for the low energy phenomenology are important. 

From the interaction Lagrangean of the fermion 
fields with the gauge bosons given in appendix B of the Ref.\cite{tuning},
we have the following terms corresponding to the fields $H_{L,R}$
\[
{\cal L}_{H} = -{q\over 2\sqrt{2}}\left[ H_{\mu,R}\sum_{\delta=1}^3 \big( 
\bar d^0_{\delta,R}\gamma^\mu d^0_{\delta,R} +
\bar u^0_{\delta,R}\gamma^\mu u^0_{\delta,R} -
\bar b^0_{\delta,R}\gamma^\mu b^0_{\delta,R} -
\bar t^0_{\delta,R}\gamma^\mu t^0_{\delta,R} \big) + \big( R\rightarrow L\big)
\right], 
\]
where $\delta$ is a color index, and the fields $u^0$,
$d^0$, $t^0$, and $b^0$ (together with $c^0$ and $s^0$) constitute a
basis for the quark fields, which must be rotated in order to get the
physical quarks, but since the couplings
of $H_{L,R}$ are not universal as can be seen from ${\cal L}_H$, FCNC
mediated by $H_{L(R)}$
(for both $Z_1$ and $Z_2$) will appear at the $m_Z$ scale, in
contradiction with the experimental bounds related to the non-existence of
low energy FCNC.

Hence the SB scheme in Ref.\cite{tuning} should be changed in order to
make the model
consistent with the low energy phenomenology. In order to do it we choose
a more economical set of
Higgs fields to implement the SB properly. We keep $\phi_3$ and
$\phi_4$ as in Ref.\cite{tuning}, but instead of $\phi_1$ and $\phi_2$ 
we use $\phi'_1$ and $\phi'_2$ both in irreps 
$Z_3\phi(\ov{15},1,15)$, with vevs along the following directions
\[\ba{ll}
\la{\phi'}_{1[\alpha,\beta]}^{[a,b]}\ra =\sqrt{3} M_H; & 
\mbox{ for } [a,b] = -[1,4] = [2,3] = [5,6]; \ \ [\alpha,\beta] = [5,6],\\
[6pt] \la{\phi'}^{[\alpha,\beta]}_{1[A,B]}\ra =\sqrt{3} M_H; & 
\mbox{ for } [a,b] = -[1,4] = [2,3] = [5,6]; \ \ [\alpha,\beta] = [4,5],\\
[6pt] \la{\phi'}_{1[a,b]}^{[A,B]}\ra = M_H; & 
\mbox{ for } [a,b],[A,B] = -[1,4] = [2,3] = - [5,6]; \\ 
\ea \]
and
\[\ba{ll}
\la{\phi'}_{2[\alpha,\beta]}^{[a,b]}\ra =\sqrt{3} M_H; & 
\mbox{ for } [a,b] = [1,2] = -[3,6] = [4,5]; \ \ [\alpha,\beta] = [5,6],\\
[6pt] \la{\phi'}^{[\alpha,\beta]}_{2[A,B]}\ra =\sqrt{3} M_H; & 
\mbox{ for } [a,b] = [1,2] = -[3,6] = [4,5]; \ \ [\alpha,\beta] = [4,5],\\
[6pt] \la{\phi'}_{2[a,b]}^{[A,B]}\ra = M_H; & 
\mbox{ for } [a,b],[A,B] = -[12] = -[3,6] = [4,5].
\ea \]
where the $\sqrt{3}$ factor is included just for convenience. 
The algebra shows that $\la\phi'_1\ra + \la\phi'_2\ra$ 
break $G$ down the LRSE model, and together with $\la\phi_3\ra$ break it
down to the SM, solving the problem discussed above.
 
\section{\uno The mass scales.}
\eq

The symmetry breaking chain is constrained by the requirement that the 
evolution of the gauge coupling constants associated with the factor
groups of 
the SM, from the $m_Z$ scale to the unification scale, 
agree with the experimental values~\cite{pdg} $sin^2 \theta_W  
(m_Z)=0.2315$,  $\alpha_{EM}^{-1}(m_Z)=127.9$, and $\alpha_3(m_Z)=0.113$. 
Then for the use of the renormalization group equations (rge)
we assume the validity of the {\it survival
hypothesis}~\cite{georgi} 
as well as the validity of the {\it extended survival hypothesis}
(which claims that when the vevs of a scalar field $\phi$
break a group $G$ down to $G_1\subset G$ at a mass scale
$M_1$, only those components of $\phi$ which acquire vevs
get a mass of the order of $M_1$, with the rest of the
components getting masses at the $G$ scale~\cite{aguila}).

When the symmetry is broken in three steps at the scales $M_R$, $M_H$ and
$m_Z$, the coupling constants satisfy, up to one loop, the rge
\be
\alpha_i^{-1}(M_Z) =
f_i\, \alpha^{-1} - b^0_i\, \ln\left({M_H\over m_Z}\right) -
b^1_i\, \ln\left({M_R\over M_H}\right),	\label{rge}
\ee
where $\alpha_i=g_i^2/4\pi$, $i =1$, 2, 3, and  $g_i$ are the gauge
coupling  constants of the $U(1)_Y$, $SU(2)_L$ and
$SU(3)_c$ subgroups of the SM respectively.  The factors $f_i$ are 
constants and define the relation at the unification scale $M$
 between $g$, the coupling constant of $[SU(6)]^3\times Z_3$  and $g_i$.
The numerical  values of these factors  $f_1 = {14/3}$, $f_2 = 3$ and $f_3 =
1$~\cite{2,tuning}, arise from the normalization conditions imposed on 
the generators of G. 

In Eq. (\ref{rge}) the beta functions $b^k_i$ are given by  
\be
b^k_i = {1\over 4\pi}\left\{ {11\over 3}C^k_i({\rm vectors}) - 
{2\over 3} C^k_i({\rm Weyl}\ {\rm fermions}) - 
{1\over 6}C^k_i({\rm scalars})\right\}, 
\ee
where $k=0,1$ and $C^k_i(\cdots)$ are the index of the representation to which
the $(\cdots)$ particles are assigned. For a complex field the value of
$C^k_i({\rm scalars})$ should be doubled. 
Also, the following relationships 
 \be 
\alpha_{EM}^{-1} \equiv \alpha_1^{-1} + \alpha_2^{-1}
\quad\mbox{ and } \quad  \tan^2\theta_W= {\alpha_1\over\alpha_2},
\ee
where $\theta_W$ is the weak mixing angle, hold at all energy scales. From
this expressions we get 
\be
{\sin^2\theta_W(m_Z)\over\alpha_{EM}(m_Z)} -  \alpha_3^{-1}(m_Z) =
\left(b_3^0-{1\over 3}b_2^0\right) 
\ln\left({M_H\over m_Z}\right) + 
\left(b_3^1-{1\over 3}b_2^1\right) 
\ln\left({M_R\over M_H}\right) \label{sin} 
\ee 
and
\be
\alpha^{-1}_{EM}(m_Z) - {23\over 3}\alpha_3^{-1}(m_Z) =
\left({23\over 3}b_3^0 -b_1^0 - b_2^0\right)
\ln\left({M_H\over m_Z}\right)  + 
\left({23\over 3}b_3^1 -b_1^1 - b_2^1\right)
\ln\left({M_R\over M_H}\right). \label{a} 
\ee
 
The analysis shows again that there is not a consistent set of solutions
for the
last two equations (the same problem was found in Refs.\cite{tuning} and 
\cite{mass}). In order to get consistent solutions we have to slightly 
modify the Higgs sector, adding
two more  Higgs fields at the scale $M_H$, $\phi^0_1$ and $\phi^0_2$ in
irreps $Z_3\phi^0_i(\ov{15},1,15), i=1,2$, with vevs in the same
directions than $\phi'_1$
and $\phi'_2$ respectively. With this new set of Higgs
fields and vevs Eqs.(\ref{sin}) and (\ref{a}) can be solved easily, 
giving $M_R\sim 10^{11}~GeVs$ and $M_H\sim 10^8~GeVs$, which are
consistent with
the seesaw neutrino mass analysis presented in Ref.~\cite{seesaw}, and 
with the bounds on low energy FCNC.


\section{\uno Concluding  remarks.}
We show with our analysis that there is not much freedom for the SB
channels of the unified model of flavors and forces based upon the local
gauge group $G$. We saw that a step of the SB implemented by vevs
of Higgs fields in irreps $Z_3\phi(\ov{n},1,n)$, with $n=15,21$,
constraint, in the $L$ sector, to break $SU(6)_L$ down only 
to some of the subgroups  
$SU(4)_L\otimes SU(2)_L$, $Sp(6)_L$, $SU(4)_L\otimes SU(2)_L$ or 
$[SU(2)_L]^3$. 

We also calculated the general form of the vevs for 
Higgs fields, in such a way that the hypercharge $Y$ of the SM does not
get broken by them, and we studied 
the possible direction for the vevs, and their breakings induced on the
different $SU(6)$ factors of $G$. 

The analysis enabled us to give an economical set of Higgs fields and vevs
which implement a SB pattern without the problems contained in the SB 
proposed in reference~\cite{tuning}, and in agreement with the
renormalization group equation analysis and the experimental data.

An important result is the existence of at least three different mass
scales, that in our case have the hierarchy
 \[M_R\sim 10^{11} GeV > M\sim 10^8~GeV>>m_z\sim 10^2 GeV,\]
with the FCNC present only at the scale $M$, in perfect agreement with the
low energy constraints.

The set of Higgs fields used (as also the set in Ref.\cite{tuning}) do not
break spontaneously the baryon number {\bf B}, which in the fundamental
irrep if $SU(6)_c$ is of the form 
${\bf B}\sim diag({1\over 3},{1\over 3},{1\over 3},0,0,0)$. 
So, the proton remains stable with the Higgs fields we have
introduced in the present analysis, and therefore we do not expect 
experimental conflicts with the upper mass scale $M_R$ calculated. 

Even though the mass hierarchy calculated here is in agreement with the
one used in the analysis of the generational seesaw mechanism, which
provide small masses to the three light neutrinos~\cite{seesaw}, we
mention that the quantitative predictions of the seesaw analysis may
depend on the particular set of Higgs fields and vevs used to break the
symmetry, specially those used for the second step of the SB because there
are not right handed neutrino mass terms from $\la\phi_3\ra$ (they come
from the Yukawa couplings between $\psi(108)$ and the scalars involved in
the second step of the SB, $\phi'_1$ and $\phi'_2$).  In this way the
changes in the scalar content of the model should affect the neutrino mass
analysis, and it should be repeated in order to check the consistence of
the previous results. Nevertheless, since the modified horizontal survival
hypothesis~\cite{2,tuning} is not violated by our new SB pattern (it is
realized by the vevs of $\phi_4$, unchanged here), and $\la\phi'_1 +
\phi'_2\ra$ produce masses of order $M_H$ for all the exotic fermions in
$\psi(6,1,\ov{6})$ and for all the vector-like particles with respect to
the LRSE model as it should be according to the survival
hypothesis~\cite{georgi}, we expect that a new seesaw analysis gives
essentially similar results.

\section*{\uno Acknowledgments.}
This work was partially supported by CONACyT, M\'exico, and COLCIENCIAS,
Colombia.

\appendix
\section*{\uno Appendix A.}
\eq
\renewcommand{\theequation}{A.\arabic{equation}}

In this appendix we present some aspects related to the branching rules of
the $SU(6)$
irreps in terms of those of their maximal subgroups. We are interested
here in the breaking of $SU(6)$ via the irreps 6, 15, 21, 35 and their
conjugates. We consider also a general $SU(6)$ group which could be
identified with whatever factor of G.

Considering all the possible
decomposition of irrep 6 of $SU(6)$ into irreps of other groups
with less dimensions (6 = 5+1,4+2 and 3+3), it is a simple matter to 
obtain the regular maximal subalgebras of $SU(6)$, they are 
$SU(5)\otimes U(1)$, $SU(4)\otimes SU(2)\otimes
U(1)$ and $SU(3)\otimes SU(3)\otimes U(1)$. Besides, $SU(6)$ also has four
special maximal subalgebras~\cite{groups} which are 
$SU(3)$, $SU(4)$, $Sp(6)$ and $SU(3)\otimes SU(2)$. 
From them the only ones containing the subgroup $SU(2)_{L(R)}$ are  
$SU(3)\otimes SU(2)$, $SU(4)\otimes SU(2)\otimes U(1)$ and $Sp(6)$. The
branching rules for $SU(3)\otimes SU(2)$ were given in Section 3.1; the
branching rules for the last two groups are
\be
\ba{rcl}
SU(6) &\rightarrow& SU(4)\otimes SU(2)\otimes U(1)\\
6 &\rightarrow& (4,1)(-1) + (1,2)(2)\\
15 &\rightarrow& (1,1)(4) + (6,1)(-2) + (4,2)(1) \\
21 &\rightarrow& (10,1)(2) + (1,3)(4) + (4,2)(1)\\
35  &\rightarrow& (1,1)(0) + (15,1)(0) + (1,3)(0) + (4,2)(-3) + (\bar 4, 2)(3);
\label{su4su2}
\ea
\ee
and
\be
\ba{rcl}
SU(6) &\rightarrow& Sp(6)\\
6 &\rightarrow& 6\\
15 &\rightarrow& 14 + 1 \\
21 &\rightarrow& 21\\
35  &\rightarrow& 14 + 21.
\ea
\ee
Therefore only the vevs of the scalar field along the singlet of a 15 may
break $SU(6)$ down to $SU(4)\otimes SU(2)\otimes U(1)$ or to $Sp(6)$
(there is no way to implement the breaking to those subgroups using irrep
21, because for the first subgroup there is not a (1,1) branching, and for
the second there is not a $Sp(6)$ singlet). From the main text we see that
irrep 21 may be used only to break $SU(6)$ down to $SU(3)\otimes SU(2)$. 

Now, from the special embedding of  $Sp(4)$ and the regular one of
$SU(2)\otimes SU(2)$ in $SU(4)$ which have the branching rules
\be
\ba{rcl}
SU(4) &\rightarrow& SU(2)\otimes SU(2)\\
4 &\rightarrow& (2,1) + (1,2)\\
6 &\rightarrow& (3,1) + (1,3)   \\
10 &\rightarrow& (3,3) + (1,1)\\
15  &\rightarrow& (3,1) + (1,3) + (3,3);
\ea\qquad
\ba{rcl}
SU(4) &\rightarrow& Sp(4)\\
4 &\rightarrow& 4\\
6 &\rightarrow& 5 + 1 \\
10 &\rightarrow& 10 \\
15  &\rightarrow& 5 + 10,
\ea
\ee
we see that there is a singlet in irrep 21 of $SU(6)$ for the group
$[SU(2)]^3$ which contains the $SU(2)_{L(R)}$ subgroup in an special
embedding. Also there is a singlet of
$Sp(4)$ in irrep 15 of $SU(6)$, and then the SB of $SU(6)$ down to
$[SU(2)]^3$ or $SU(4)\otimes SU(2)$ using a single Higgs (in one 
step) is always possible.

The breaking of $SU(6)$ to any of the other maximal subalgebras  
necessarily
break the $SU(2)_L$ group structure, therefore there may be paths allowed
only for vevs with indices
in the $c$ and $R$ spaces. 

The branching rules for the other two regular maximal subalgebras are 
\be
\ba{rcl}
SU(6) &\rightarrow& SU(5)\otimes U(1)\\
6 &\rightarrow& 5(1) + 1(-5)\\
15 &\rightarrow& 10(2) + 5(-4) \\
21 &\rightarrow& 15(2) + 5(-4) + 1(-10) \\
35  &\rightarrow& 24(0) + 1(0) + 5(6) + \bar{5}(-6);
\ea
\ee
and
\be
\ba{rcl}
SU(6) &\rightarrow& SU(3)\otimes SU(3)\otimes U(1)\\
6 &\rightarrow& (3,1)(1) + (1,3)(-1)\\
15 &\rightarrow& (\ov{3},1)(2) + (1,\ov{3})(-2) + (3,3)(0)\\
21 &\rightarrow& (6,1)(2) + (1,6)(-2) + (3,3)(0)\\
35  &\rightarrow& (8,1)(0) + (1,8)(0) + (1,1)(0) + (3,\ov{3})(2) +
(\ov{3},3)(-2).
\ea
\ee
So, the breaking down to $SU(5)$ is possible only by the vevs in irrep 21,
while there is not a lower dimension scalar field able to break $SU(6)$
down to its regular subgroup $SU(3)\otimes SU(3)$.

For the other special subalgebras of $SU(6)$ we have the branching rules 
\be
\ba{rcl}
SU(6) &\rightarrow& SU(4)\\
6 &\rightarrow& 6\\
15 &\rightarrow& 15 \\
21 &\rightarrow& 20 + 1\\
35  &\rightarrow& 15 + 20;
\ea\qquad
\ba{rcl}
SU(6) &\rightarrow& SU(3)\\
6 &\rightarrow& 6\\
15 &\rightarrow& 15 \\
21 &\rightarrow& 15+ 6\\
35  &\rightarrow& 8 + 27.
\ea
\ee
Again, the breaking down to the special subgroup $SU(4)$ may be
implemented
only via vevs in irrep 21, while neither 15 nor 21 could do the
breaking down to the special $SU(3)$ subgroup.

To conclude, notice that the breaking of $SU(6)$  down to the non maximal
subalgebra $SU(4)\times U(1)$ is possible by vevs along the irreps $(1,3)$
of $SU(4)\otimes SU(2)$ in irrep 21, and also that the breaking of $SU(6)$
down to the special maximal subgroup $SU(3)\otimes SU(2)$
is not possible via irreps 15, 21 or 35 as it can be seen from
Eq.(\ref{su2su3}). (A further analysis shows that it is
possible only via
irrep 105 in $SU(6)$).

\end{document}